\documentclass[journal,comsoc]{IEEEtran}
\usepackage[T1]{fontenc}
\usepackage{cite}
\usepackage{amsmath}
\usepackage[cmintegrals]{newtxmath}
\usepackage[hyphens]{url}
\usepackage{graphicx}
\usepackage{algorithm}
\usepackage{algpseudocode}
\usepackage{subfigure}
\usepackage{diagbox}
\usepackage{multirow}
\usepackage{authblk}
\usepackage{array}
\newcolumntype{P}[1]{>{\centering\arraybackslash}p{#1}}
\newcolumntype{M}[1]{>{\centering\arraybackslash}m{#1}}
\ifCLASSINFOpdf
\else
\fi
\begin{document}
%
\title{An Efficient Fog-Assisted Unstable Sensor Detection Scheme with Privacy Preserved}
%
%
%

\author[1]{Shuo Chen}
\author[2]{Rongxing Lu}
\author[1]{Jie Zhang}

\affil[1]{School of Computer Science and Engineering, Nanyang Technological University, Singapore}
\affil[2]{Faculty of Computer Science, University of New Brunswick, Canada \authorcr Email: {\{schen, zhangj\}}@ntu.edu.sg, rxlu@ieee.org}

\renewcommand\Authands{ and }
\maketitle

\begin{abstract}
The Internet of Thing (IoT) has been a hot topic in both research community and industry. It is anticipated that in future IoT, an enormous number of sensors will collect the physical information every moment to enable the control center making better decisions to improve the quality of service (QoS). However, the sensors maybe faulty and thus generate inaccurate data which would compromise the decision making. To guarantee the QoS, the system should be able to detect faulty sensors so as to eliminate the damages of inaccurate data. Various faulty sensor detection mechanisms have been developed in the context of wireless sensor network (WSN). Some of them are only fit for WSN while the others would bring a communication burden to control center. To detect the faulty sensors for general IoT applications and save the communication resource at the same time, an efficient faulty sensor detection scheme is proposed in this paper. The proposed scheme takes advantage of fog computing to save the computation and communication resource of control center. To preserve the privacy of sensor data, the Paillier Cryptosystem is adopted in the fog computing. The batch verification technique is applied to achieve efficient authentication. The performance analyses are presented to demonstrate that the proposed detection scheme is able to conserve the communication resource of control center and achieve a high true positive ratio while maintaining an acceptable false positive ratio. The scheme could also withstand various security attacks and preserve data privacy.
\end{abstract}

\begin{IEEEkeywords}
faulty sensor detection, fog computing, privacy-preserving, security, Internet of Thing.
\end{IEEEkeywords}

%
\IEEEpeerreviewmaketitle

\section{Introduction}
\label{sec1}
%
%
%
%
\IEEEPARstart{W}{ith} the prosperous developing of communication and computation technologies, the Internet of Things, which allows the physical objects to be accessed and controlled remotely through the network infrastructure and thus more tightly integrates the physical world and cyber world, is no longer a fancy nowadays. The big advantage of IoT is that with the data analysis on the huge amount of information collected from the physical world, the control center is capable of making more accurate and optimal decisions which would produce considerable benefits. It is estimated that the global IoT market will be 14.4 trillion dollars by 2022 \cite{Bradley14}, the potential economic impact of IoT would be 3.9 to 11.1 trillion dollars by 2025 \cite{McKinsey2015}, and the number of devices connected to the Internet would be about 50 billion by 2020 \cite{Cisco2014}. Since the control center makes decisions based on the data analysis, the requirement of data accuracy should be high. However, in reality, the data analysis could suffer from the interferences caused by some physical factors. Specifically, the sensors collecting physical information maybe faulty due to device aging, battery depletion or some unexpected environmental influences, which could result in inaccurate data. There are several sensor fault types defined in \cite{ni2009sensor}. In this paper, the sensor fault type we focus on is \emph{high noise or variance} which may be caused by hardware failure or low batteries \cite{ni2009sensor}. For this type of faulty sensor, its readings will have deviations from actual physical data due to unusually high amount of noise. Denote the actual physical data at time $t$ as $data(t)$ and sensor reading at time $t$ as $data^{\ast}(t)$. A smart sensor is defined as faulty sensor if $data^{\ast}(t) = data(t) \cdot (1 + \delta)$ where $\delta$ is a random variable with normal distribution $N(0, \alpha^2)$ and the deviation measurement $\alpha^2$ is larger than the inherent white noise of sensor. The sensing data with high amount of noise would not only waste the computation and communication resource but also compromise the data analysis result of control center. It is impractical to check the physical condition of all sensors regularly because the amount of sensors is too large and many sensors are deployed in the unattended or uncontrollable environment. In the state of the art, there are several approaches for detecting the faulty sensors. Some of these methods are based on majority voting, which compare the sensor's data with its neighbor sensors to determine whether the sensor is normal \cite{chen2006distributed,lee2008fault}. However, the state of the neighbor sensors would affect the detection accuracy, i.e. if the majority of the neighbors are faulty, a normal sensor would be recognized as faulty. Moreover, the size of neighbour sensor set also affect the detection effectiveness. Those methods require that there are several sensors performing the same task, which is suitable for WSN while maybe infeasible for other IoT applications. Thus, these kinds of methods are inadequate for the general IoT scenarios. The other kinds of detection schemes use control center to analyse the time series sensor data by bayesian belief network, machine learning, wavelet analysis, principal component analysis, etc \cite{mehranbod2005method,warriach2013fault,yang2015multi,dunia1996identification}. Due to the huge amount of sensors, the data transmission from sensors to control center would consume considerable communication resource and may lead to network delay. 

\begin{figure}[htbp]
	\centering
	\includegraphics[width=0.5\textwidth]{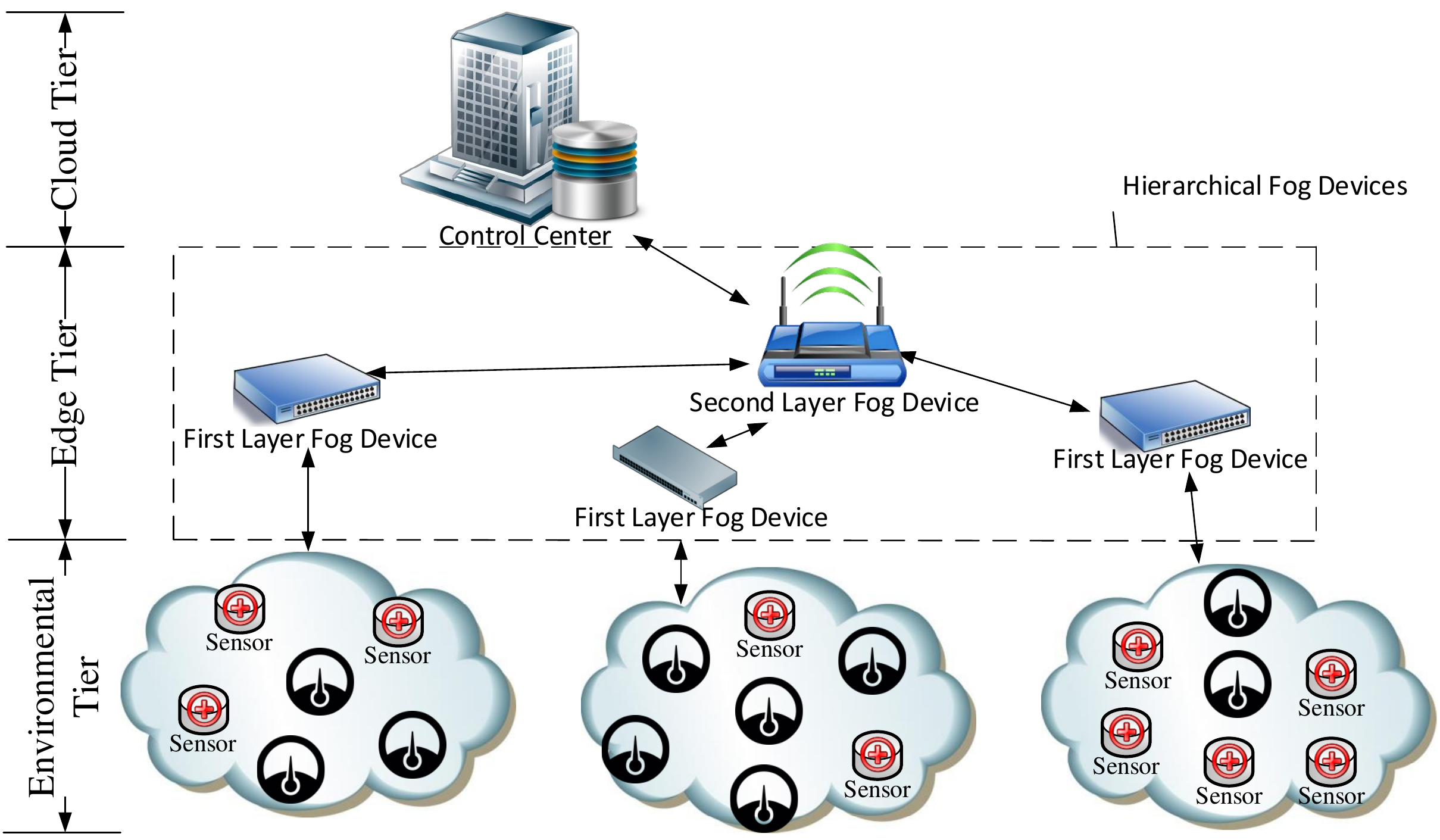}
	\caption{Architecture of Fog Computing}\label{fig1}
\end{figure}

To decrease the sensor data volume transmitted to control center for ensuring the network quality of service, a new technique called fog computing, which is proposed by Cisco \cite{bonomi2012fog}, is suitable to be applied. The main idea of fog computing is to provide storage, computing and networking services between end devices and control center. The fog devices which are in close proximity to end devices normally possess considerable storage and computation resource. With the equipped resource, the fog devices could process the collected data locally so as to ease the burden of control center. The architecture of fog computing is shown in Fig. \ref{fig1}.


There are three tiers in the fog computing architecture: environmental tier, edge tier and cloud tier. In the environmental tier, there are billions of smart sensors collecting heterogeneous information of physical world, e.g. medical sensors in eHealth and smart meters in smart grid. The data obtained by the smart sensors will be transmitted to the edge tier. The distribution of fog devices in edge tier are hierarchical which is a characteristic inherited from the traditional network architecture. For example, the switchers of a local area network could function as the first layer fog devices and the gateway which manages those switchers could serve as the second layer fog device. The fog devices in edge tier will perform the fog processing to deal with the received sensor data locally and then send the results to the control center in cloud tier. 
Owing to the processing of fog devices, the volume of data sent to control center could be reduced to a large extent. Since the fog devices are spread in a highly distributed environment, it is impractical for the government or an institution to provide and maintain all those devices. Therefore, it is reasonable to assume that the fog devices would be supplied by third parties. 

Under the context of fog computing, one could migrate the abovementioned time series data analysis from control center to fog devices for faulty sensor detection, i.e. the fog devices analyse the received data with the traditional methods and report the detection result to control center. However, in many IoT applications, the data collected from the environment would be considered as privacy by the data owners, e.g. the vital signs in eHealth and the power usage in smart grid. Performing the traditional analysis with fog devices is infeasible if the privacy is a primary concern from the perspective of data owners because what the traditional methods analyse are plaintexts and the third parties which control the fog devices may not be trustworthy. Therefore, how to take advantage of fog computing to locally detect faulty sensors in a privacy-preserving way is a challenging issue.


 
In this paper, we propose an efficient fog-assisted faulty sensor detection scheme with privacy-preserving, which could identify the faulty sensors locally so as to save the communication resource of control center. Since the sensor fault type this paper focuses on is \emph{high noise or variance}, inspired by \cite{wang2015optimal}, the scatter matrix is adopted to measure the degree of sensor data deviation. The homomorphic encryption technique called Paillier encryption \cite{paillier1999public} is applied to ensure the fog devices compute the scatter matrix in a privacy-preserving way. Utilizing the scatter matrix, the fog devices could calculate the dispersion measure and then identify the faulty sensors. The main contributions of this paper are three-fold.

\begin{itemize}
	\item First, to identify faulty sensors, we propose an efficient fog-assisted privacy-preserving detection scheme. Since the detection procedures are performed in the fog devices, it could reduce the communication resource consumption of control center significantly. Moreover, for each sensor, even its data is multidimensional, the corresponding processed data transmitted from the first layer fog device to the second layer fog device would still be a single Paillier ciphertext. The scatter matrix could be derived from the ciphertext efficiently by the second layer fog device. The feature could conserve the local network resource.
	\item Second, the proposed scheme utilizes Paillier Cryptosystem to preserve the data privacy. Each data sample is encrypted as a Paillier ciphertext before being sent to fog devices. Utilizing the feature of Paillier Cryptosystem, the first layer fog device could perform algebraic operations on sensor data in the encrypted form, which would not leak the sensor data content. The second layer fog device could only obtain the aggregated data from the first layer fog device to compute the scatter matrix. Thus, the sensor data privacy could be preserved during the fog processing.
	\item Third, our scheme achieves the data confidentiality, authenticity and integrity. With the application of Paillier Cryptosystem and Boneh-Lynn-Shacham (BLS) short signature technique \cite{boneh2004short}, the messages exchanged among the environmental tier, edge tier and cloud tier could withstand the eavesdropping attack, replay attack and man-in-the-middle attack.
\end{itemize}

The reminder of this paper is organized as follow. In Section \ref{sec2}, the system model, security requirements and design goals are described. The preliminaries of our scheme are introduced in Section \ref{sec3}. In Section \ref{sec4}, the proposed detection scheme is presented in details. The security analysis and performance evaluation are discussed in Section \ref{sec5} and \ref{sec6}, respectively. In Section \ref{sec7}, the related work discussion is presented. Finally, our work is concluded in Section \ref{sec8}.

\section{System Model, Security Requirements and Design Goals}
\label{sec2}

In this section, we describe the system model, formalize the security requirement and identify the design goals on privacy-preserving faulty sensor detection.
\subsection{System Model}
\begin{figure}[htbp]
	\centering
	\includegraphics[width=0.5\textwidth]{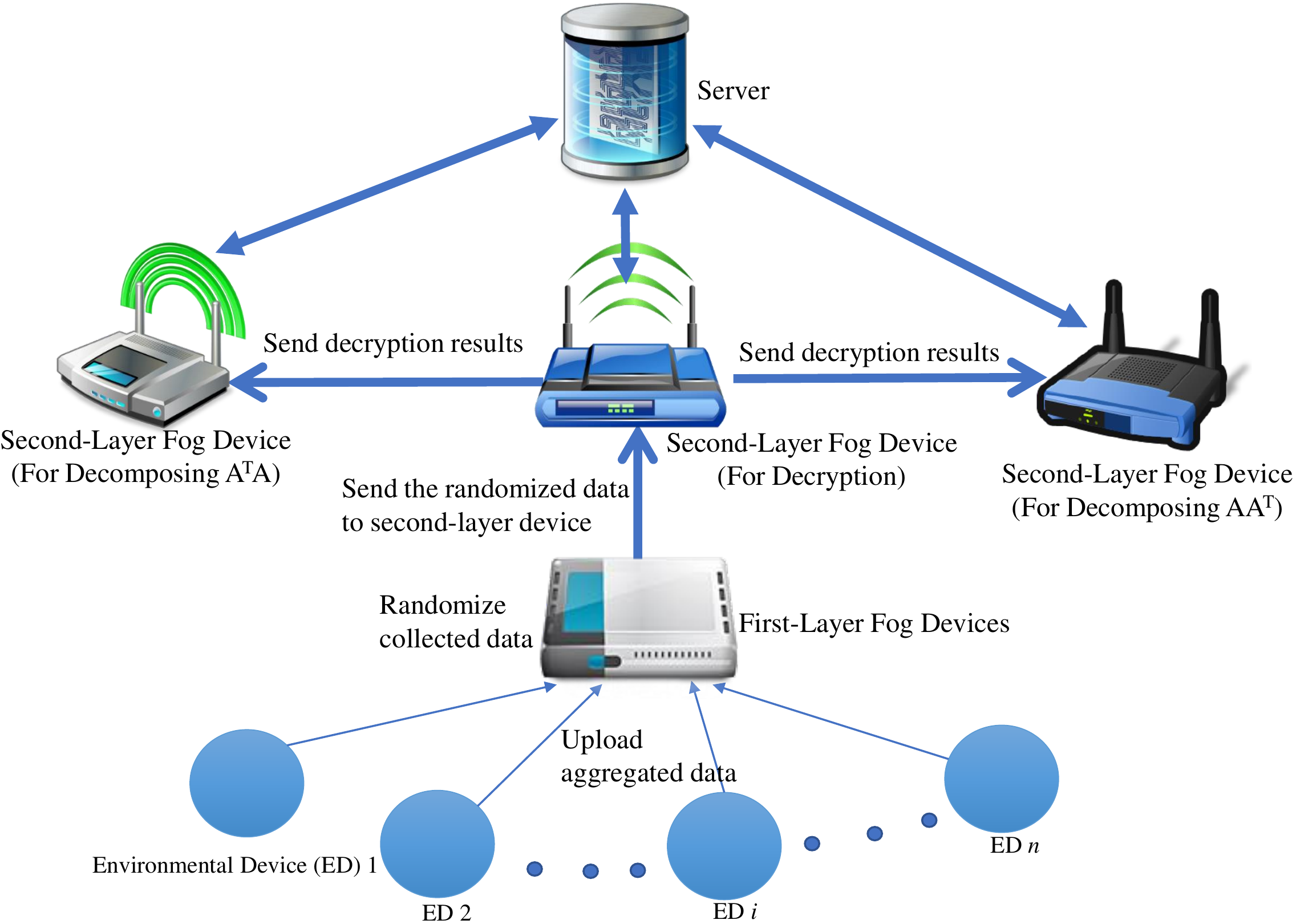}
	\caption{System Model}\label{fig2}
\end{figure}

In this work, we mainly focus on how to utilize the fog computing to detect faulty sensor locally and with privacy-preserving. Specifically, there are four entities in the system model, namely a control center, a first layer fog device, a second layer fog device and a smart sensor to be detected as shown in Fig. \ref{fig2}.

\textbf{Control Center (CC)}: CC is a fully trustable entity locating in the cloud tier. It is responsible for initializing the whole system and distributing key materials to others.

\textbf{First Layer Fog Device (FD)}: FD is the fog device which communicates with the smart sensor directly. FD receives and processes the data from smart sensor, and uploads the processed data to the second layer fog device. FD is assumed to possess a considerable amount of computation and storage resource which could be used to do cryptography and data aggregation operations.

\textbf{Second Layer Fog Device (SD)}: SD is the fog device which communicates with FD. Upon receiving the data from FD, SD would perform further processing on the data for the purpose of detecting faulty sensor. Similar to FD, SD is assumed to be capable of performing cryptography operations.

\textbf{Smart Sensor (SS)}: SS collects environmental information which is formed as multidimensional data periodically. The fog devices determine whether SS is faulty through analysing the data uploaded by SS. SS is assumed to have little storage and could only do cryptography operations.

In our model, CC first initializes the whole system and distributes the key materials to other entities accordingly. After initialization, SS keeps sensing the physical information. Each time SS collects a physical data sample, it will encrypt and sign it, and send the sample to FD. FD stores the received samples for a certain period of time. When the number of received samples meets the preset value, FD would carry out its processing procedures to aggregate the data samples into a single Paillier ciphertext which contains the full information needed to compute the scatter matrix. Upon receiving the aggregated data from FD, SD will decrypt the Paillier ciphertext and further compute the scatter matrix and dispersion measure. Comparing the dispersion measure with a certain threshold, SD could determine whether SS is faulty. Then, SD reports the detection result to CC. Finally, if SS is recognized as faulty, CC could advise the system manager to check the condition of SS manually and fix/replace SS if necessary. This could improve the quality of service which is of paramount importance for IoT.

\emph{Communication model.} Considering the scope of local region, WiFi technology is utilized to achieve the communication among SS, FD and SD. For the communication between SD and CC, the channel is composed of wired/wireless links with high bandwidth due to the long distance. Note that although with high bandwidth channel, the communication efficiency between cloud tier and edge tier is still a challenging issue if the enormous amount of sensor data from environmental tier is relayed to CC directly.


\subsection{Security Requirements}
Security is fundamental for the effectiveness of faulty sensor detection. In our security model, CC and SS are assumed to be trustable. The hierarchical fog devices, i.e. FD and SD, are assumed to be honest-but-curious which means they will follow the processing procedures faithfully while being curious about the sensor data contained in the messages. In addition, FD and SD are assumed not to collude with each other. Besides the four entities in the system model, there exists an adversary $\mathcal{A}$ whose capabilities are assumed as: \emph{1)} $\mathcal{A}$ could eavesdrop the messages exchanged among the cloud tier, edge tier and environmental tier. \emph{2)} $\mathcal{A}$ could replay and modify the eavesdropped messages to launch some active attacks. Based on the above assumptions, we have the following security requirements:
\begin{itemize}
	\item \emph{Confidentiality.} The sensor data should not be disclosed to fog devices and adversary $\mathcal{A}$. Specifically, for FD and SD, even they process the collected data, they could not learn anything about the situation of monitored object; for the adversary $\mathcal{A}$, even it could eavesdrop the messages exchanged among legitimate entities, it could not gain any knowledge of monitored object.
	\item \emph{Authenticity and Integrity.} The messages transmitted in the system should be authenticated to ensure they are generated by legal entities and have not been altered. This is to protect the system from the active attacks launched by the adversary $\mathcal{A}$. Any malicious actions from the adversary $\mathcal{A}$ should be detected.
\end{itemize}

\subsection{Design Goals}
According to the aforementioned system model and security requirements, the design goal is to develop an efficient privacy-preserving faulty sensor detection scheme with the assistance of fog computing. Specifically, the proposed scheme should achieve the following objectives.
\begin{itemize}
	\item \emph{The faulty sensors should be identified with the proposed detection scheme.} Since the faulty data would waste the system resource and influence the decision making of CC, the proposed scheme should achieve a high true positive ratio and maintain a relatively low false positive ratio.
	
	\item \emph{The communication resource should be conserved in the proposed scheme.} The communication effectiveness between cloud tier and edge tier could be affected if all the sensor data is transmitted to CC for analysis. The proposed scheme should manage to reduce the overall data volume sent to CC.
	
	\item \emph{The confidentiality, authenticity and data integrity should be guaranteed in the proposed scheme.} All the messages containing the information of monitored object need to be protected in case of eavesdropping and modifying. The fog processing in FD and SD should not leak the privacy of monitored object.	 
\end{itemize}

\section{Preliminaries}
\label{sec3}
In this section, the Paillier Cryptosystem \cite{paillier1999public}, bilinear pairing technique \cite{boneh2001identity} and scatter matrix \cite{wang2015optimal} which are the basis of the proposed scheme are reviewed.
\subsection{Paillier Cryptosystem}
The Paillier Cryptosystem enables the addition and multiplication operations on plaintext through the specific linear algebraic manipulation conducted on the ciphertext. This property is extensively desired in many privacy-preserving applications \cite{sang2009privacy, zhong2007privacy, lu2012eppa}. In this paper, this feature allows FD to process the sensor data in encrypted form without leaking the data content. The Paillier Cryptosystem comprises three phases: key generation, encryption and decryption.
\begin{itemize}
	\item \emph{Key Generation: }Given one security parameter $\kappa$,generate two large prime numbers $p$, $q$, where $|p|=|q| = \kappa$. Then compute the $\mathcal{RSA}$ modulus $n  = p q$, $\lambda = lcm(p-1, q-1)$ and choose a generator $g \in \mathbb{Z}_{n^2}^{\ast}$. Define the function $L(u) = \frac{u - 1}{n}$ and calculate $\mu = (L(g^{\lambda} \bmod n^2))^{-1} \bmod n$. Then \emph{\textbf{PK}} $= (n, g)$ is published as the public key and \emph{\textbf{SK}} $= (\lambda, \mu)$ is kept secret as the corresponding private key.
	\item \emph{Encryption: }Given a message $m \in \mathbb{Z}_n$, randomly choose a number $r \in \mathbb{Z}_n^{\ast}$, the ciphertext could be calculated as $c = E(m, r) = g^m \cdot r^n \bmod n^2$.
	\item \emph{Decryption: }Given the ciphertext $c \in \mathbb{Z}_{n^2}^{\ast}$, the corresponding plaintext could be recovered as $m = D(c) = L(c^{\lambda}  \bmod n^2)\cdot \mu \bmod n$. Note that, the Paillier Cryptosystem is provably secure against chosen plaintext attack, and the correctness and security can be referred to \cite{paillier1999public}.
\end{itemize}
%

\subsection{Scatter Matrix}
The scatter matrix is used to evaluate the data set divergence. It is widely applied in pattern classification \cite{duda2012pattern} \cite{szeliski2010computer}, feature selection \cite{van2005classification} and principal component analysis \cite{jolliffe2002principal} for multiple areas, e.g. data mining and computer vision. Given $N$ samples of $l$-dimensional data, represented as the $l$-by-$N$ matrix, $\mathbb{X}$ = [$\vec{x}_1$, $\vec{x}_2$, $\cdots$, $\vec{x}_N$], the data mean is
\begin{equation}
	\overline{x} = \frac{1}{N}\sum_{i = 1}^{N}\vec{x}_i
\end{equation}
where $\vec{x}_i$ is the $i$th column of $\mathbb{X}$. The scatter matrix $\mathbb{M}$ is an $l$-by-$l$ positive semi-definite matrix computed as
\begin{equation}
	\mathbb{M} = \frac{1}{N} {\sum_{i = 1}^{N}(\vec{x}_i - \overline{x})(\vec{x}_i - \overline{x})^T}
\end{equation}
where $T$ denotes matrix transpose. As mentioned in \cite{wang2015optimal}, obtaining the scatter matrix, one could find its eigenvalues through decomposing the scatter matrix into a diagonal matrix.
\begin{equation}
	\mathbb{M} = Pdiag(\lambda_1,\cdots,\lambda_l)P^{-1}
\end{equation}
The $i$th eigenvalue $\lambda_i$ measures the variance of the $i$th data dimension. The dispersion measure is defined as the multiplication of nonzero eigenvalues, and a smaller dispersion measure value indicates a more concentrated set. In this paper, we utilize this feature to detect faulty sensors. In specific, for each sensor, we collect $N$ data samples from it in a certain period of time, and compute the scatter matrix and dispersion measure accordingly. Since the samples are from the same sensor, the data values of the same dimension should be highly correlated. If the sensor is normal, the variance of collected data samples should be relatively small which leads to a small dispersion measure. In other words, if the dispersion measure is larger than a certain threshold, it means the sensor is under faulty state. Note that the physical data used for detection purpose should be collected when the condition of monitored object is steady since what the scatter matrix measures is the dispersion degree of data set. If the data generated by the monitored object is originally unsteady, the variance of data set would be inevitably larger than normal, and a normal sensor maybe falsely recognized as faulty. For simplicity, we assume the monitored object is under steady condition when the detection scheme is being carried out. 
\section{Proposed Scheme}
\label{sec4}
In this section, the proposed efficient fog-assisted faulty sensor detection scheme with privacy-preserving is presented in details. The scheme is composed of four phases: system initialization, sensor data collection, fog aggregation and fog analysis.
\subsection{System Initialization}
CC is the trustable entity which bootstraps the whole system. Specifically, given the security parameters $\kappa$, CC calculates the public key for Paillier Cryptosystem \emph{\textbf{PK}}: $(n = p q, g)$, and the corresponding private key \emph{\textbf{SK}}: $(\lambda, \mu)$, where $p$, $q$ are two large primes with $|p| = |q| = \kappa$. Also, given the security parameters $\kappa_1$, CC runs $\mathcal{G}$en$(\kappa_1)$ to generate $(q_1, P, \mathbb{G}, \mathbb{G}_T, e)$ for bilinear pairing. Set the required number of data samples for analysis as $N$. Assume the sensor data is $l$-dimensional and the value range for each data dimension is [$0$, $d$] even if the sensor is faulty, where $d$ is a constant. Then, CC chooses two superincreasing sequence $\vec{\mathbf{a}} = (a_1 = 1, a_2,\cdots, a_l)$ and $\vec{\mathbf{b}} = (b_1 = 1, b_2,\cdots, b_N)$ such that $\sum_{j = 1}^{i-1}b_j \cdot 2 \cdot N \cdot d < b_i$ for $i = 2,\cdots, N$, $\sum_{j=1}^{i-1}a_j \cdot {\sum_{k = 1}^{N}b_k \cdot 2 \cdot N \cdot d} < a_i$ for $i = 2,\cdots, l$ and $\sum_{i=1}^{l}a_i \cdot {\sum_{k = 1}^{N}b_k \cdot 2 \cdot N \cdot d} < n$. Also, CC chooses one secure cryptographic hash function $H$ where $H: \{0, 1\}^{\ast} \rightarrow \mathbb{G}$. For SD, FD and SS, CC randomly chooses numbers $x_s$, $x_f$ and $x_u$ $\in \mathbb{Z}_{q_1}^{\ast}$ and computes $Y_s = x_sP$, $Y_f = x_fP$ and $Y_u = x_uP$ respectively. Finally, CC publishes the public parameters as $\{q_1, P, \mathbb{G}, \mathbb{G}_T, e, n, g, Y_s, Y_f, Y_u, H, \vec{\mathbf{a}}, \vec{\mathbf{b}}\}$, sends the $(\lambda, \mu, x_s)$ to SD as secret, and distributes the secrets $x_f$ and $x_u$ to FD and SS respectively.

\subsection{Sensor Data Collection}
In environmental tier, SS keeps monitoring the physical situation and reporting the $l$-dimensional data sample $(d_1,\cdots, d_l)^T$ frequently. 
For the $i$th data sample $(d_{1i},\cdots, d_{li})^T$, SS performs the following steps:
\begin{itemize}
	\item \emph{Step-1. }Utilize the superincreasing $\vec{\mathbf{a}}$ to compute
	\begin{equation}
	m_i = a_1d_{1i}+a_2d_{2i}+\cdots+a_ld_{li}
	\end{equation}
	\item \emph{Step-2. }Choose a random number $r_i \in \mathbb{Z}_n^{\ast}$ and compute
	\begin{equation}
	C_i = g^{m_i} \cdot r_i^n \bmod n^2
	\end{equation}
	\item \emph{Step-3. }Use the private key $x_u$ to compute the signature $\sigma_i$ as
	\begin{equation}
	\sigma_i = x_uH(C_i||SS||TS_i)
	\end{equation}
	where $TS_i$ is the current time stamp.
	\item \emph{Step-4. }Send the encrypted and signed $i$th data sample $C_i||SS||TS_i||\sigma_i$ to FD.
\end{itemize}
\subsection{Fog Aggregation}
When receiving a data sample, FD will check its time stamp and store the sample if the time stamp is acceptable. Upon receiving total $N$ data samples from SS, FD first performs batch verification to verify their signatures, i.e. check whether $e(P, \sum_{i=1}^{N}\sigma_i) \stackrel{?}{=} \prod_{i=1}^{N}e(Y_u, H(C_i||SS||TS_i))$. If the equation does hold, it means the samples are valid. The correctness is as follow.
	\begin{equation}
	\begin{split}
		e(P, \sum_{i=1}^{N}\sigma_i) &= e(P, \sum_{i=1}^{N}x_uH(C_i||SS||TS_i))\\
		&= \prod_{i=1}^{N}e(P, x_uH(C_i||SS||TS_i))\\
		&= \prod_{i=1}^{N}e(Y_u, H(C_i||SS||TS_i))\\
	\end{split}
	\end{equation}
After the validity checking, FD performs the following privacy-preserving steps to aggregate the $N$ data samples from SS.
\begin{itemize}
	\item \emph{Step-1. }Aggregate the $N$ encrypted samples as
	\begin{equation}
		\begin{split}
			C &= \prod_{i=1}^{N}C_i \bmod n^2\\
			&= \prod_{i=1}^{N}g^{m_i} \cdot r_i^n \bmod n^2\\
			&= g^{\sum_{i=1}^{N}m_i} \cdot \prod_{i=1}^{N}r_i^n \bmod n^2\\
			&= g^{a_1\sum_{i=1}^{N}d_{1i}+\cdots+a_l\sum_{i=1}^{N}d_{li}} \cdot \prod_{i=1}^{N}r_i^n \bmod n^2\\
		\end{split}
	\end{equation}
	\item \emph{Step-2. }Calculate $C_a = g^{\sum_{i=1}^{l}a_i \cdot d} \bmod n^2$. For each $C_i, i = 1,\cdots, N$,  perform
	\begin{equation}
	\begin{split}
	CD_i &= \frac{(C_i \cdot C_a)^N}{C} \bmod n^2\\
	&= \frac{(g^{a_1d_{1i}+\cdots+a_ld_{li}+\sum_{i=1}^{l}a_i \cdot d} \cdot r_i^n \bmod n^2)^N}{C} \bmod n^2\\
	&= g^{a_1 \cdot [N(d_{1i}+d)-\sum_{k=1}^{N}d_{1k}]+\cdots+a_l \cdot [N(d_{li}+d)-\sum_{k=1}^{N}d_{lk}]} \cdot r_{di}^n\\
	& \bmod n^2\\
	&= g^{\sum_{j=1}^{l}a_j \cdot [N(d_{ji}+d)-\sum_{k=1}^{N}d_{jk}]} \cdot r_{di}^n \bmod n^2\\
	\end{split}
	\end{equation}
	where $r_{di} = r_i^N \cdot (\prod_{k=1}^{N}r_k)^{-1} \bmod n^2$. Note that multiplying $C_a$ with  $C_i$ is to make $N(d_{ji}+d)-\sum_{k=1}^{N}d_{jk} > 0$, for $j = 1, \cdots, l$. This ensures that SD could utilize the superincreasing sequence to recover the information needed for computing scatter matrix. The detailed explanation is presented in the \emph{correctness of Algorithm \ref{Alg:Cal}} below.
	\item \emph{Step-3. } Aggregate all the $CD_i, i = 1,\cdots, N$, as
	\begin{equation}
		R_f = \prod_{i=1}^{N}CD_i^{b_i} \bmod n^2
	\end{equation}
	The $R_f$ is the aggregation result of the data samples from SS.
	\item \emph{Step-4. }Use the private key $x_f$ to compute the signature $\sigma_f = x_fH(R_f||SS||FD||TS_f)$, where $TS_f$ is the current time stamp.
	\item \emph{Step-5. }Send the aggregation result $R_f||SS||FD||TS_f||\sigma_f$ to SD.
\end{itemize}
\subsection{Fog Analysis}
Upon receiving $R_f||SS||FD||TS_f||\sigma_f$, SD first checks $e(P, \sigma_f) \stackrel{?}{=} e(Y_f, H(R_f||SS||FD||TS_f))$ to verify the time stamp and signature. Then SD will perform the following steps to analyse the aggregation result $R_f$ for SS and send report to CC. The $R_f$ is implicitly formed by
\begin{equation}
	\begin{split}
	R_f &= \prod_{i=1}^{N}CD_i^{b_i} \bmod n^2\\
	&= \prod_{i=1}^{N}g^{b_i \cdot \sum_{j=1}^{l}a_j \cdot [N(d_{ji}+d)-\sum_{k=1}^{N}d_{jk}]} \cdot r_R^n \bmod n^2\\
	&= g^{\sum_{i=1}^{N}b_i \cdot \sum_{j=1}^{l}a_j \cdot [N(d_{ji}+d)-\sum_{k=1}^{N}d_{jk}]} \cdot r_R^n \bmod n^2\\
	&= g^{\sum_{j=1}^{l}a_j \cdot \sum_{i=1}^{N}b_i \cdot [N(d_{ji}+d)-\sum_{k=1}^{N}d_{jk}]} \cdot r_R^n \bmod n^2\\
	\end{split}
\end{equation}
where $r_R = \prod_{i=1}^{N} (r_{di})^{b_i}$.
\begin{itemize}
	\item \emph{Step-1. }	Utilize the secret key $(\lambda, \mu)$ to decrypt the $R_f$ and get the plaintext of aggregation result
	\begin{equation}
		M_f = \sum_{j=1}^{l}a_j \cdot \sum_{i=1}^{N}b_i \cdot [N(d_{ji}+d)-\sum_{k=1}^{N}d_{jk}] \quad \bmod n
	\end{equation} 
	\item \emph{Step-2. }Through running Algorithm \ref{Alg:Cal}, calculate the scatter matrix from $M_f$.

\begin{algorithm}
	\footnotesize
	\caption{Calculate the scatter matrix from aggregated data}\label{Alg:Cal}
	\begin{algorithmic}[1]
		\Procedure {Calculate the scatter matrix}{}
			
		\noindent\textbf{Input:} $\vec{\mathbf{a}}=(a_1=1, \cdots, a_l)$, $\vec{\mathbf{b}}=(b_1=1, \cdots, b_N)$  and $M_f$	
		\noindent\textbf{Output:}Scatter Matrix of SS
		\State{Let $\vec{Temp_a} = (t_{a1}, \cdots, t_{al})^T$ and $\mathbb{T}emp_b = [\vec{t_{b1}}, \cdots, \vec{t_{bN}}]$ where $\vec{t_{bi}} = (t_{b1i}, \cdots, t_{bli})^T, i = 1, \cdots, N$.}		
		\State{Set $X_l = M_f$}
		\For{$j = l$ to $2$}
		\State{$X_{j-1} = X_j \bmod a_j$}
		\State{$t_{aj} = \frac{X_j - X_{j-1}}{a_j} = \sum_{i = 1}^{N}b_i \cdot [N(d_{ji}+d)-\sum_{k=1}^{N}d_{jk}]$}
		
		\EndFor
		\State{$t_{a1} = X_1  = \sum_{i = 1}^{N}b_i \cdot [N(d_{1i}+d)-\sum_{k=1}^{N}d_{1k}]$}
		\For{$j =1$ to $l$}
		\State{Set $X_N = t_{aj}$}
		\For{$i=N$ to $2$}
		\State{$X_{i-1} = X_i \bmod b_i$}
		\State{$t_{bji} = \frac{\frac{X_i - X_{i-1}}{b_i}}{N} - d = d_{ji} - \frac{1}{N} \sum_{k=1}^{N}d_{jk}$}
		\EndFor
		\State{$t_{bj1} = \frac{X_1}{N} - d = d_{j1} - \frac{1}{N} \sum_{k=1}^{N}d_{jk}$}
		\EndFor
		\State{\textbf{return} $\frac{1}{N}\sum_{i=1}^{N}\vec{t_{bi}} \cdot \vec{t_{bi}}^T$}
		\EndProcedure
	\end{algorithmic}
\end{algorithm}

\item \emph{Step-3. }Compute the dispersion measure of SS. Then determine whether SS is a faulty sensor through comparing its dispersion measure with a preset threshold. Let report be $1||SS$ if SS is normal and $0||SS$ otherwise. Use the private key $x_s$ to compute the signature $\sigma_s = x_sH(report||SD||TS_s)$,	where $TS_s$ is the current time stamp.
\item \emph{Step-4. }Send the report $report||SD||TS_s||\sigma_s$ to CC. Note that if the report is considered as confidential, SD could also encrypt it before sending it to CC.
\end{itemize}

\emph{\textbf{The correctness of Algorithm~\ref{Alg:Cal}}}. In Algorithm~\ref{Alg:Cal}, $X_l = M_f = a_1\sum_{i=1}^{N}b_i \cdot [N(d_{1i}+d)-\sum_{k=1}^{N}d_{1k}] + \cdots + a_{l-1}\sum_{i=1}^{N}b_i \cdot [N(d_{(l-1)i}+d)-\sum_{k=1}^{N}d_{(l-1)k}] + a_l\sum_{i=1}^{N}b_i \cdot [N(d_{li}+d)-\sum_{k=1}^{N}d_{lk}] \bmod n$. Since the data value for each dimension is in the range of [$0$, $d$], we have
\begin{equation}\label{}
\begin{split}
&a_1\sum_{i=1}^{N}b_i \cdot [N(d_{1i}+d)-\sum_{k=1}^{N}d_{1k}] + \cdots \\
& \qquad \qquad + a_{l-1}\sum_{i=1}^{N}b_i \cdot [N(d_{(l-1)i}+d)-\sum_{k=1}^{N}d_{(l-1)k}]\\
<& a_1\sum_{i=1}^{N}b_i \cdot N \cdot 2 \cdot d + \cdots + a_{l-1}\sum_{i=1}^{N}b_i \cdot N \cdot 2 \cdot d \\
=& \sum_{j=1}^{l-1} a_j \sum_{i=1}^{N}b_i \cdot N \cdot 2 \cdot d  < a_l \\
\end{split}
\end{equation}
Therefore, $X_{l-1} = X_l \bmod a_l = a_1\sum_{i=1}^{N}b_i \cdot [N(d_{1i}+d)-\sum_{k=1}^{N}d_{1k}] + \cdots + a_{l-1}\sum_{i=1}^{N}b_i \cdot [N(d_{(l-1)i}+d)-\sum_{k=1}^{N}d_{(l-1)k}]$, and
\begin{equation}
	\begin{split}
	\frac{X_{l} - X_{l-1}}{a_l} &= \frac{a_l\sum_{i=1}^{N}b_i \cdot [N(d_{li}+d)-\sum_{k=1}^{N}d_{lk}]}{a_l}\\ 
	&= \sum_{i=1}^{N}b_i \cdot [N(d_{li}+d)-\sum_{k=1}^{N}d_{lk}] = t_{al}\\
	\end{split}
\end{equation}
Similarly, it can be proved that  $t_{aj} = \sum_{i = 1}^{N}b_i \cdot [N(d_{ji}+d)-\sum_{k=1}^{N}d_{jk}]$, for $j = 1, \cdots, l-1$. Obtaining $t_{aj}$, for $j = 1, \cdots,l$, we have $X_N = t_{aj} = b_1 \cdot [N(d_{j1}+d)-\sum_{k=1}^{N}d_{jk}] + \cdots + b_N \cdot [N(d_{jN}+d)-\sum_{k=1}^{N}d_{jk}]$. Since $0 \le d_{ji} \le d$, for $i = 1, \cdots, N$ and $j = 1, \cdots, l$, there is
\begin{equation}\label{}
\begin{split}
&b_1 \cdot [N(d_{j1}+d)-\sum_{k=1}^{N}d_{jk}] + \cdots \\
& \qquad \qquad + b_{N-1} \cdot [N(d_{j(N-1)}+d)-\sum_{k=1}^{N}d_{jk}] \\
<&b_1 \cdot N \cdot 2 \cdot d + \cdots + b_{N-1} \cdot N \cdot 2 \cdot d \\
=& \sum_{i=1}^{N-1} b_i \cdot N \cdot 2 \cdot d  < b_N \\
\end{split}
\end{equation}
Therefore, $X_{N-1} = X_N \bmod b_N = b_1 \cdot [N(d_{j1}+d)-\sum_{k=1}^{N}d_{jk}] + \cdots + b_{N-1} \cdot [N(d_{j(N-1)}+d)-\sum_{k=1}^{N}d_{jk}]$, and
\begin{equation}
\begin{split}
\frac{\frac{X_{N} - X_{N-1}}{b_N}}{N} - d
&= d_{jN}-\frac{1}{N}\sum_{k=1}^{N}d_{jk} = t_{bjN}\\ 
\end{split}
\end{equation}
Similarly, it can be proved that  $t_{bji} = d_{ji}-\frac{1}{N}\sum_{k=1}^{N}d_{jk}$, for $i = 1, \cdots, N-1$ and $j = 1, \cdots, l$. Obtaining all the $t_{bji}$, the scatter matrix could be calculated as  $\frac{1}{N}\sum_{i=1}^{N}\vec{t_{bi}} \cdot \vec{t_{bi}}^T$. As a result, the correctness of Algorithm~\ref{Alg:Cal} is shown.

Note that in the \emph{Step-2} of fog aggregation, $C_a$ is used to multiply $C_i$ to guarantee $N(d_{ji}+d)-\sum_{k=1}^{N}d_{jk} > 0$, for $j = 1, \cdots, l$. If we did not multiply $C_a$ with $C_i$, $M_f$ would become $\sum_{j=1}^{l}a_j \cdot \sum_{i=1}^{N}b_i \cdot [N d_{ji}-\sum_{k=1}^{N}d_{jk}] \bmod n$ and $\sum_{i=1}^{N}b_i \cdot [N d_{ji}-\sum_{k=1}^{N}d_{jk}]$ maybe negative. The negative $\sum_{i=1}^{N}b_i \cdot [N d_{ji}-\sum_{k=1}^{N}d_{jk}]$ could make the analysis result wrong. For example, if $d_{lN} = 0$ and $d_{lk} = d$, for $k = 1, \cdots, N-1$, then
\begin{equation}
\begin{split}
\sum_{i=1}^{N}b_i \cdot [N d_{li}-\sum_{k=1}^{N}d_{lk}] &= \sum_{i=1}^{N-1}b_i\cdot d + b_N[-(N-1)d]\\
& < b_N-b_N[(N-1)d] < 0 
\end{split}
\end{equation}
Assuming $\sum_{i=1}^{N}b_i \cdot [N d_{ji}-\sum_{k=1}^{N}d_{jk}] > 0$ for $j = 1, \cdots, l-1$, then $t_{al} = \frac{1}{a_l}(M_f - M_f \bmod a_l) = \sum_{i=1}^{N}b_i \cdot [N d_{li}-\sum_{k=1}^{N}d_{lk}] \bmod n$ and $t_{blN} = \frac{1}{N}\frac{t_{al} - t_{al} \bmod b_N}{b_N}$. Due to the modular operation, $t_{al}$ would become $n - \sum_{i=1}^{N}b_i \cdot [N d_{li}-\sum_{k=1}^{N}d_{lk}]$, which is a positive number. This leads to $t_{blN}$ being positive if $t_{al} \ge b_N$, or 0 if $t_{al} < b_N$. However, the correct $t_{blN}$ should be $d_{lN}-\frac{1}{N}\sum_{k=1}^{N}d_{lk}$ which is negative. So it is necessary to multiply $C_a$ with $C_i$ to ensure the information is correctly recovered.

\section{Security Analysis}
\label{sec5}
In this section, the security properties of the proposed detection scheme are analysed. Specifically, the analysis focuses on how the proposed scheme could achieve the aforementioned security requirements in section \ref{sec2}.
\begin{itemize}
	\item \emph{Sensor data privacy is preserved in the proposed detection scheme. }In the proposed scheme, each sensor data sample $(d_1, \cdots, d_l)^T$ is encrypted as $g^{\sum_{j=1}^{l}a_jd_j} \cdot r^n \bmod n^2$, which is a ciphertext of Paillier Cryptosystem. During fog aggregation, what FD does is performing modular power, division and multiplication operations on the collection of ciphertexts. When the fog aggregation result is decrypted in SD, the plaintext is formed as $M = \sum_{j=1}^{l}a_j \cdot \sum_{i=1}^{N}b_i \cdot [N(d_{ji}+d)-\sum_{k=1}^{N}d_{jk}]$. With the published $\vec{\mathbf{a}}$ and $\vec{\mathbf{b}}$, $d_{ji}-\frac{1}{N}\sum_{k=1}^{N}d_{jk}, i = 1,\cdots,N; j = 1, \cdots,l$ could be derived from $M$. Since Paillier Cryptosystem is semantic secure against the chosen plaintext attack and only SD, which does not collude with FD, possesses the secret key $(\lambda, \mu)$, $m = \sum_{j=1}^{l}a_jd_j$ is semantic secure to FD and the adversary $\mathcal{A}$. Thus the confidentiality of sensor data $(d_1, \cdots, d_l)^T$ is guaranteed under fog aggregation and eavesdropping. In addition, even SD recovers each $d_{ji}-\frac{1}{N}\sum_{k=1}^{N}d_{jk}$, it could not learn anything about the $d_{ji}$ because it does not know the value of data mean $\frac{1}{N}\sum_{k=1}^{N}d_{jk}$. Therefore, the proposed scheme could preserve the data privacy.
	
	\item \emph{The authenticity and integrity of messages are achieved in the proposed detection scheme. }In the proposed scheme, all transmitted messages are signed by the legal parties with the BLS short signature technique \cite{boneh2004short}. The source authenticity and message integrity could be guaranteed because the BLS short signature is provably secure under the CDH problem in random oracle model \cite{bellare1993random}. The time stamp contained in the messages could prevent replay attack. Any man-in-the-middle attack, e.g. alteration of time stamp and message content, could be noticed since the signature validity checking would fail. Therefore, our proposed scheme is capable of detecting the malicious activities from adversary $\mathcal{A}$.
\end{itemize}

Based on the above analysis, it is obvious that our proposed detection scheme achieves the privacy-preserving property while guaranteeing the authenticity and integrity at the same time, which fulfils the security requirements in section \ref{sec2}. 
\section{Performance Evaluation}
\label{sec6}

%
%

In this section, we evaluate the performance of the proposed detection scheme in terms of capacity, effectiveness, computation complexity and communication overhead. Let $l$, $d$ and $N$ denote the data dimension, maximum data value, required number of data samples from SS for detection.
\subsection{Capacity}
\begin{figure*}[!htbp]
	\begin{center}
		\subfigure[Ciphertext space is 2048 bits]
		{\label{n2048}\includegraphics[width=0.48\textwidth]{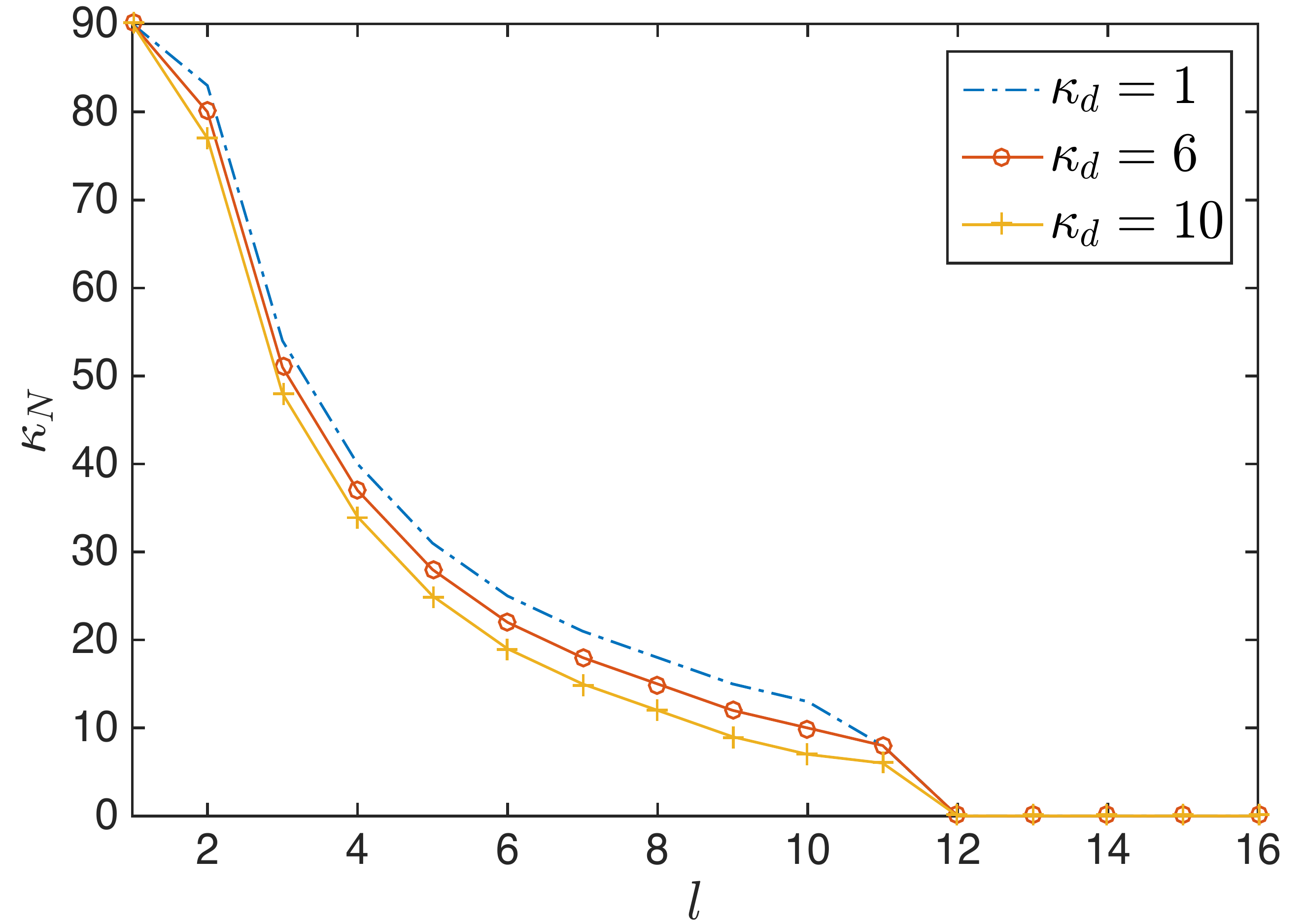}}
		\subfigure[Ciphertext space is 4096 bits]
		{\label{n4096}\includegraphics[width=0.48\textwidth]{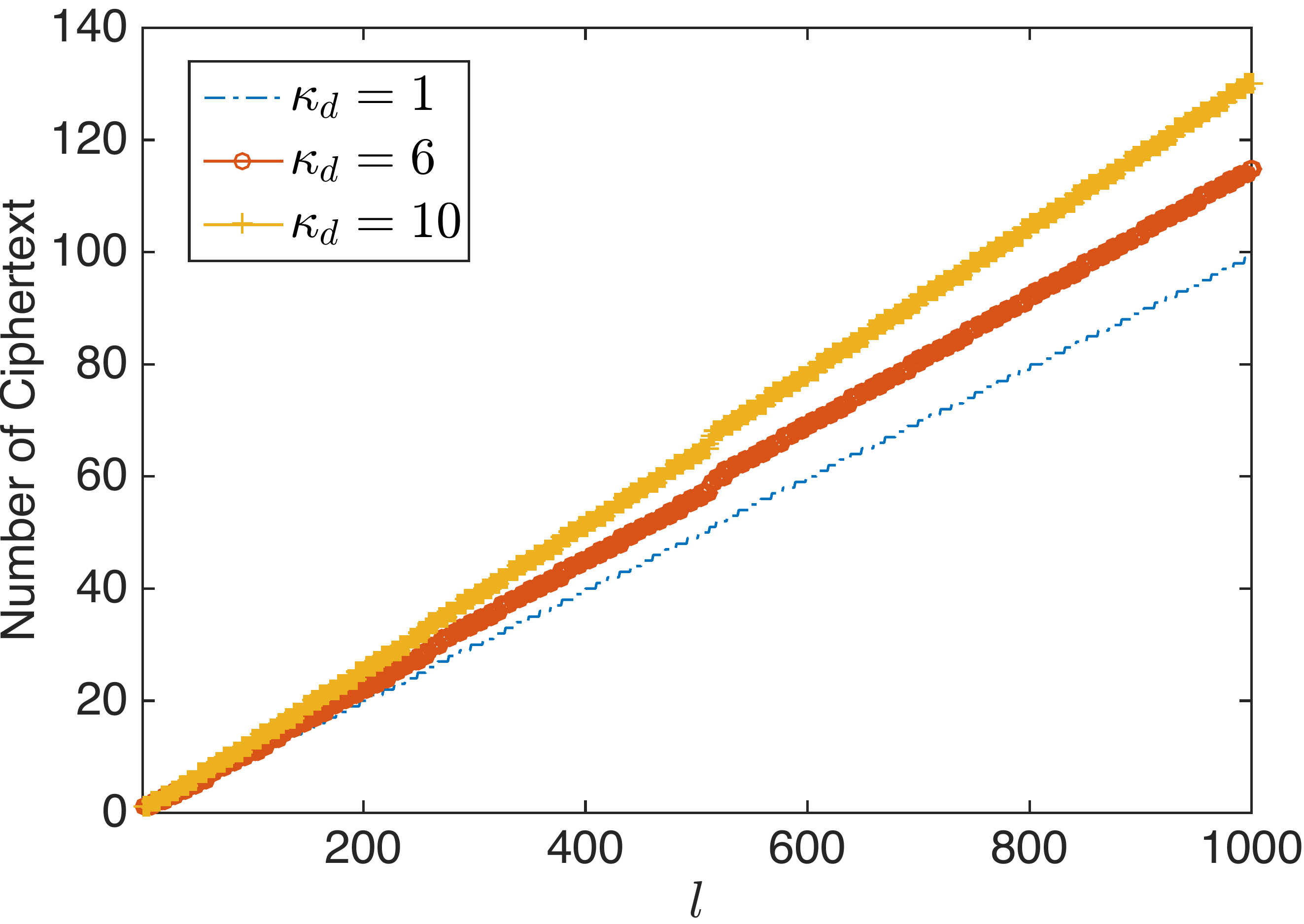}}
	\end{center}
	\caption{Capacity of proposed scheme}
	\label{fig3}
\end{figure*}

In the proposed scheme, $N$ data samples from SS are aggregated as one Paillier ciphertext for detection. To guarantee the aggregated data $M_f$ could be recovered correctly from the ciphertext through decryption, the constraint $\sum_{i=1}^{l}a_i \cdot {\sum_{k = 1}^{N}b_k \cdot 2 \cdot N \cdot d} < n$ must be fulfilled. However, the superincreasing sequence $\vec{\mathbf{a}}$ and $\vec{\mathbf{b}}$ also need to meet the constraints: $\sum_{j = 1}^{i-1}b_j \cdot 2 \cdot N \cdot d < b_i$ for $i = 2,\cdots, N$ and $\sum_{j=1}^{i-1}a_j \cdot {\sum_{k = 1}^{N}b_k \cdot 2 \cdot N \cdot d} < a_i$ for $i = 2,\cdots, l$. As the value of $N$, $l$ and $d$ increases, the value of the elements in $\vec{\mathbf{a}}$ and $\vec{\mathbf{b}}$ will grow very fast, which would cause it impossible to meet the constraint $\sum_{i=1}^{l}a_i \cdot {\sum_{k = 1}^{N}b_k \cdot 2 \cdot N \cdot d} < n$. Therefore, in this part, given different data dimension $l$ and data range [$0$, $d$], we evaluate the maximum number of data samples could be aggregated into one Paillier ciphertext in the proposed scheme. We implement the proposed detection scheme with Java programming language and the evaluation result is shown in Fig. \ref{fig3}.

From Fig. \ref{fig3}, it could be shown that the maximum number of data samples FD could aggregate is very sensitive to the data dimension while the impact of data value range is not that significant. To collect more data samples for detection, one could choose a larger ciphertext space.

\subsection{Effectiveness}
In this part, we utilize real physical data to evaluate the effectiveness of the proposed scheme in terms of true positive ratio (TPR) and false positive ratio (FPR). Let $N_{uu}$ denotes the number of faulty data sets detected as faulty, $N_{tu}$ denotes the total number of faulty data sets, $N_{nu}$ denotes the number of normal data sets detected as faulty and $N_{tn}$ denotes the total number of normal data sets. The definition of TPR and FPR are as follow.

\begin{equation}
\begin{split}
TPR = \frac{N_{uu}}{N_{tu}} \quad , \quad FPR = \frac{N_{nu}}{N_{tn}}\\
\end{split}
\end{equation}  
The real data is 2-dimensional ECG signal selected from the MIT-BIH Arrhythmia Database \cite{physiobank2000physionet,moody2001impact}. Specifically, the data file we choose is the 100.dat file and there are 1300000 data samples inside. There are two phases in the evaluation which are training phase and testing phase.

\emph{Training Phase} In this phase, we select the first half of data samples in the 100.dat file as the training samples and utilize them to choose the proper thresholds to guarantee the TPR is higher than 95\% under different deviations of data. Specifically, for a certain deviation, among the 650000 training samples, we randomly choose 10000 data sets as training sets each of which contains 10 data samples from a continuous time period, i.e. N = 10. Then we randomly choose 2000 data sets from the training sets to inject deviations, i.e. $N_{tu} = 2000$ and $N_{tn} = 8000$. For each data sample [$d_1$, $d_2$] from the selected 2000 data sets, the deviations are injected as $d_1^{\ast} = d_1 + \left \lfloor (d_1 \cdot \delta_1)\right \rfloor$ and $d_2^{\ast} = d_2 + \left \lfloor (d_2 \cdot \delta_2)\right \rfloor$ where $\delta_1$ and $\delta_2$ are random variables with normal distribution $N(0, \alpha^2)$, the value of $\alpha^2$ is the measurement of deviation and the $\left \lfloor Y \right \rfloor$ means the largest integer less than or equal to $Y$. After the deviations are injected, the scatter matrix and dispersion measure are computed for each training data set. Then we compare the dispersion measures with a threshold $Th$. For each dispersion measure, if its value is larger than $Th$ and the corresponding data set is injected deviation before, $N_{uu}$ increases by 1. After all the comparisons, TPR is calculated as $\frac{N_{uu}}{2000}$. If TPR $\ge$ $95\%$, the threshold value is selected, otherwise we decrease the value of $Th$ and repeat the comparisons. The selected threshold values under three different deviation measurements are shown in Table \ref{thvalue}. From the Table \ref{thvalue}, it could be shown that a smaller threshold could help to detect a tinier data deviation. Thus, if the requirement for data accuracy is very high, a small threshold should be selected.

\begin{table}[htbp]
	\renewcommand\arraystretch{1.3}
	\centering
	\caption{Selected Threshold Values}
	\label{thvalue}
	\begin{tabular}{|p{0.05\textwidth}|p{0.1\textwidth}|p{0.1\textwidth}|p{0.1\textwidth}|}
		\hline
		$\alpha^2$ &$1\%$ &$5\%$ &$10\%$\\
		\hline
		$Th$ & $3.8 \times 10^4$ & $10^7$ &$1.5 \times 10^8$\\
		\hline
	\end{tabular}	
\end{table}

\emph{Testing Phase} In this phase, the second half of data samples in the 100.dat file is chosen as the testing samples. For simplicity, we assume the system could tolerate $5\%$ data deviation brought by faulty sensors. Then the testing samples and the previously selected threshold $10^7$ are utilized to evaluate the TPR and FPR the proposed scheme could achieve with different data set sizes and under different data deviations which are higher than or less than $5\%$. In specific, for a certain data set size $N$ and deviation measurement $\alpha^2$, we select 10000 testing data sets each of which contains $N$ data samples from a continuous time period. Similar as the previous phase, 2000 data sets are randomly chosen from the testing data sets and injected deviations. Then we compare the dispersion measures with the selected threshold $Th$. If a dispersion measure is larger than $Th$ and the corresponding data set is injected deviation before, $N_{uu}$ increases by 1, otherwise $N_{nu}$ increases by 1. Finally, the TPR is calculated as $\frac{N_{uu}}{2000}$ and FPR is calculated as $\frac{N_{nu}}{8000}$. The simulation results with different $\alpha^2$ and $N$ are show in Table \ref{th1}. 
According to Table \ref{th1}, it could be shown that when the deviation measurement is higher than $5\%$, the proposed scheme could always achieve a TPR higher than $95\%$ with the preset threshold while maintaining a relatively small FPR. Moreover, collecting more data samples for detection normally could achieve a higher TPR while bringing a larger FPR as the price.


\begin{table}[htbp]
	\renewcommand\arraystretch{1.3}
	\centering
	\caption{Performance of proposed scheme}
	\label{th1}
	\begin{tabular}{|p{0.05\textwidth}|p{0.05\textwidth}|p{0.05\textwidth}|p{0.05\textwidth}|p{0.05\textwidth}|p{0.05\textwidth}|}
		\hline
		\multicolumn{6}{|c|}{$Th = 10^7$}\\
		\hline
		\multicolumn{2}{|c|}{\diagbox[width=9em,trim=l]{$\alpha^2$}{$N$}} & 10 & 15 & 20 & 25\\
		\hline
		\multirow{2}{*}{$4\%$} & TPR & 0.8115  & 0.9175 & 0.9610 & 0.9850 \\\cline{2-6}
		& FPR & 0.0808& 0.1041 & 0.1190 & 0.1432\\
		\hline
		\multirow{2}{*}{$4.5\%$} & TPR & 0.9240 & 0.9740 & 0.9895 & 0.9940 \\\cline{2-6}
		& FPR & 0.0843 & 0.1046 & 0.1149 & 0.1420\\
		\hline
		\multirow{2}{*}{$5\%$} & TPR & 0.9560  & 0.9875 & 0.9950 & 0.9970 \\\cline{2-6}
		& FPR & 0.0815 & 0.1051 & 0.1154 & 0.1409\\
		\hline
		\multirow{2}{*}{$5.5\%$} & TPR & 0.9815   & 0.9935 & 0.9955 & 0.9945 \\\cline{2-6}
		& FPR & 0.0818 & 0.1043 & 0.1189 & 0.1426\\
		\hline
		\multirow{2}{*}{$6\%$} & TPR & 0.9865  & 0.9950 & 0.9950  & 0.9970 \\\cline{2-6}
		& FPR & 0.0811 & 0.1029 & 0.1177 & 0.1411\\
		\hline		
	\end{tabular}	
\end{table}


According to the above analysis, it is shown that with a proper threshold selected based on the maximum allowed data deviation, the proposed scheme could achieve a high TPR and keep the FPR relatively small. In addition, the proposed scheme is also very flexible, i.e. when there are spare resource, the proposed scheme could collect more data samples for analysis to achieve a higher TPR.

\subsection{Computation Complexity}
In the proposed detection scheme, when SS generates an encrypted and signed data sample $C_i||SS||TS_i||\sigma_i$, it needs to perform $2$ exponentiation operation in $\mathbb{Z}_{n^2}$ to produce $C_i$, and $1$ multiplication operation in $\mathbb{G}$ to generate $\sigma_i$. Upon receiving $N$ data samples from SS, FD needs $(N+1)$ pairing operations for signature verification. During the fog aggregation, the cost for aggregating $C_i$ as $C$ is negligible since the multiplication in $\mathbb{Z}_{n^2}$ is considered as negligible compared to exponentiation and pairing operations. Note that the $C_a = g^{\sum_{i=1}^{l}a_i \cdot d} \bmod n^2$ could be calculated in advance and utilized repeatedly, so the cost for generating $C_a$ could be ignored. To calculate $CD_i$ and aggregate $CD_i$ as $R_f$, FD requires $2N$ exponentiation operations in $\mathbb{Z}_{n^2}$. Additionally, $1$ multiplication operation in $\mathbb{G}$ is required for FD to generate the signature for aggregation result. As for SD, it has to perform $2$ pairing operations every time it receives an aggregation result from FD. After verification, SD needs to perform 1 exponentiation operation in $\mathbb{Z}_{n^2}$ to decrypt the aggregation result. Obtaining the plaintext of aggregation result, the calculating of scatter matrix and dispersion measure would be performed on plaintext, of which computation cost is negligible compared to other operations. Finally, to guarantee the authenticity and integrity of its report, SD would further perform $1$ multiplication operation in $\mathbb{G}$ to generate the signature. Upon receiving the report from SD, CC needs to perform $1$ pairing operation to verify the signature. Let $C_e$, $C_m$ and $C_p$ denote the resource consumptions of an exponentiation operation in $\mathbb{Z}_{n^2}$, a multiplication operation in $\mathbb{G}$ and a pairing operation respectively. To determine whether SS is faulty, the total required computation resource for SS, FD, SD and CC will be $N(2C_e+C_m)$, $[(N+1)C_p+2NC_e+C_m]$, $(2C_p+C_e+C_m)$ and $C_p$ respectively in the proposed scheme. It could be seen that most of the computation resource is consumed in fog devices rather than CC which means the fog devices ease the computation burden of CC through processing data locally.

%
\subsection{Communication Overhead}
In this part, the consumption of communication resource from local region to CC is evaluated. For the comparison with the proposed fog-assisted scheme, we consider the traditional approach (denoted by TRAD). In TRAD approach, SS encrypts its data sample with AES-128 technique as $C_i = E_{AES}(d_{1i}||\cdots||d_{li})$ and generates the signature as in the proposed scheme. Then SS sends the encrypted and signed data to FD. Upon receiving $N$ data sample from SS, FD performs batch verification to verify their signatures. Then FD concatenates the $N$ samples as $C_1||\cdots||C_N$ and generates a signature for it. After that, FD sends the concatenated message and signature to SD. Since the concatenated message contains the sensor data, SD would not be granted the ability to decrypt it. Therefore, SD relays the message directly to CC. After decrypting the message, CC would utilize the traditional methods mentioned in Section \ref{sec1} to determine whether SS is faulty. For the encryption in TRAD, we assume one AES-128 block is enough to encapsulate the $l$-dimensional data. 
For the signature generation, we assume a $160$-bits $\mathbb{G}$ are chosen.

\begin{figure}[htbp]
	\centering
	\includegraphics[width=0.5\textwidth]{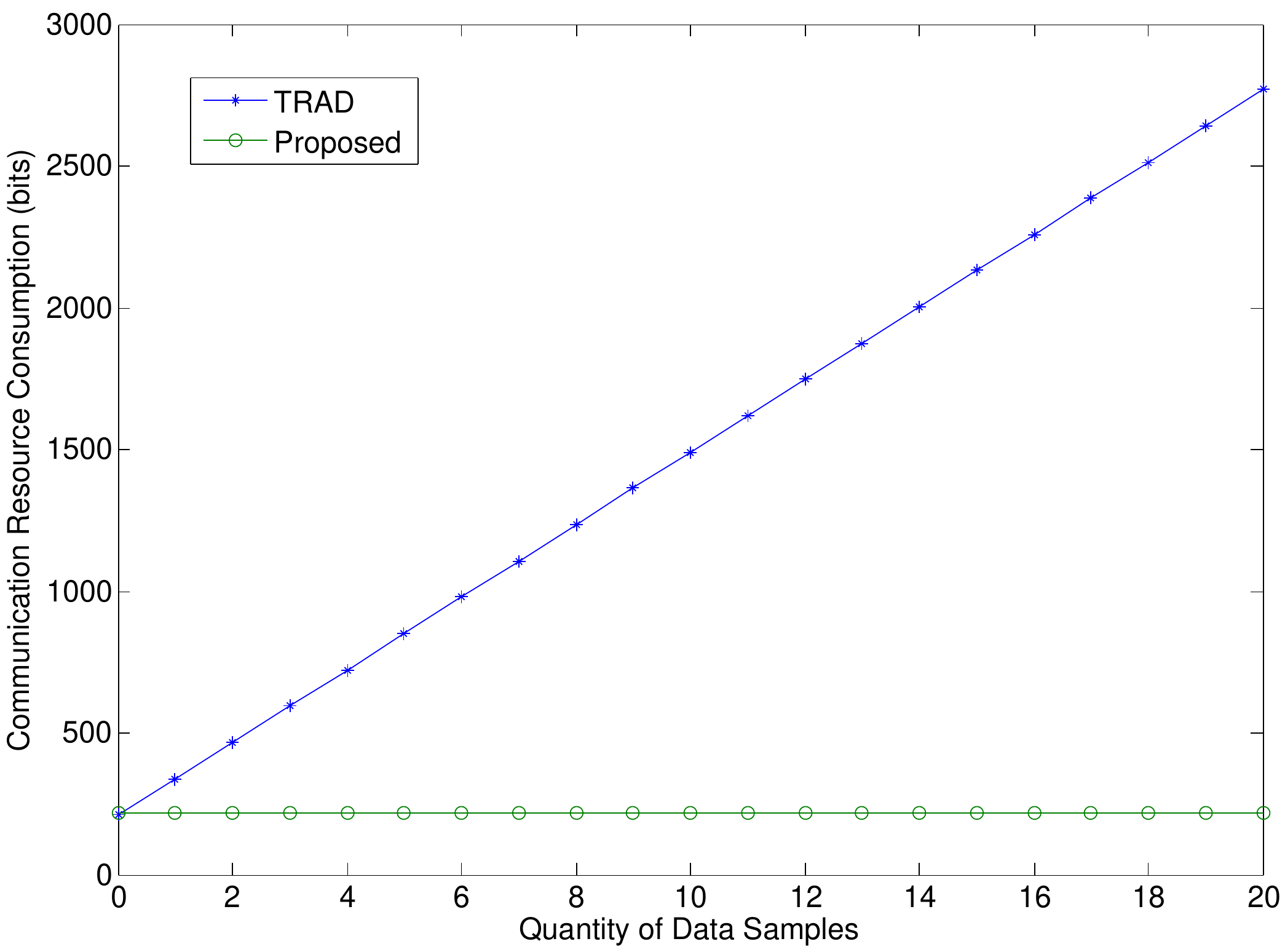}
	\caption{Comparison of Communication Resource Consumption}\label{fig4}
\end{figure}

In TRAD approach, since FD concatenates $N$ data samples and SD just relays the message, the size of concatenated message sent to CC is $S_{TRAD} = (N \cdot 128+|SS|+|TS|+160)$ bits. In the proposed scheme, since SD is able to perform the detection locally, the size of data sent to CC is $S_{proposed} = (|report|+|SD|+|TS|+160)$ bits . We plot the communication overhead of both schemes in terms of $N$, as shown in Fig. \ref{fig4}, where we assume the $(|SS|+|TS|)$ are 50 bits and $(|report|+|SD|+|TS|)$ are 60 bits. It is shown that the proposed scheme exceeds the TRAD approach in terms of the communication resource consumption. With the increasing of $N$, the advantage of our scheme becomes more significant. Note that just for one smart sensor, the proposed scheme could reduce that much communication cost. Considering the huge amount of sensors spread in IoT, the total communication bandwidth saved by the proposed scheme would be substantial.

From the above analysis, the proposed detection scheme manages to relieve the communication load of CC remarkably with the assistance of fog computing, which would contribute to realizing the future IoT.

\section{Related Works}
\label{sec7}
In this section, the previous research works related to this paper are reviewed.

\emph{Faulty Sensor Detection Schemes} Chen et al. \cite{chen2006distributed} proposed a distributed fault detection scheme based on majority voting for WSN. Their scheme compares the measurement of the sensor with that of its neighbours at time $t$. If the measurement difference is higher than a threshold, the scheme compares the measurement difference at time $t$ with that at time $t-1$. If more than half of the neighbours have a considerably different value and the measurement differences change over the time greatly, the sensor is determined as faulty. Their simulation results demonstrate that almost all the sensors could be diagnosed correctly with large network size. They also claim that in realistic situation, not enough neighbours would cause the incorrect diagnosis. Lee et al. \cite{lee2008fault} proposed another distributed faulty sensor detection scheme which considers the time delay when comparing the sensor reading with its neighbours. They employ a sliding window to eliminate the impact of transmission delay. The simulation shows that their scheme could identify the faults with high accuracy. Since their scheme is also based on majority voting, they claim that the existence of malicious nodes may compromise the detection results. These distributed schemes are all designed for WSN environment and could efficiently detect the sensor faults to guarantee the overall system performance. However, in many IoT applications, the number of sensors performing the same task are usually limited. Even to guarantee the reliability, the system may just deploy 2 or 3 sensors for the same object. Therefore, the majority voting based mechanisms are not feasible for general IoT scenarios. The proposed detection scheme of this paper only analyzes the data from the single sensor to be detected, which is suitable for the general IoT applications.

Mehranbod et al. \cite{mehranbod2005method} proposed a bayesian belief network based sensor fault detection scheme. The scheme is designed for the scenario in which one node (parent node) has causal influence on the other (child node). The correlation between the parent and child node is represented by the conditional probability and the probability absolute difference is the measurement used to detect faults. Warriach et al. \cite{warriach2013fault} proposed a machine learning based approach to detect sensor faults in WSN. The hidden markov models is adopted to capture the dynamics of a fault-free environment and dynamics of faulty data. Then the structural analysis of the models is performed to detect the sensor measurement faults. Yang et al. \cite{yang2015multi} proposed a multi-level wavelet Shannon entropy based scheme to locate the single-sensor fault. The scheme firstly chooses the appropriate wavelet base for signal analysis according to the criterion of maximum energy-to-Shannon entropy ratio. Then the sensor fault is located based on the wavelet time Shannon entropy and wavelet time-energy Shannon entropy. Dunia et al. \cite{dunia1996identification} proposed a faulty sensor identification scheme utilizing the principal component analysis. The scheme uses the principal component model to capture the measurement correlations and reconstruct each variable. Then the status of each sensor is identified based on a proposed sensor validity index. All the above methods require a control center to analyse the plaintext of the sensor data, which would consume a large amount of communication resource of core networks. The fog-assisted faulty sensor detection scheme proposed in this paper is capable of detecting the faulty sensor locally and thus conserve the network resource.

\emph{Fog Computing Applications} The fog computing has attracted more and more attentions recently. Cao et al. \cite{cao2015distributed} proposed a real-time fall detection system utilizing fog computing. The detection task is split between the fog devices and the server in their system to automatically detect the falls in a timely manner. Dubey et al. \cite{dubey2015fog} proposed a service-oriented architecture for fog computing in telehealth applications. The low power embedded computers in the architecture carry out data analytics on collected raw data and transmit unique patterns to server. Their implementation shows that the data volume transmitted to server is reduced and hence saving the transmission power. Aazam et al. \cite{aazam2014fog} presented the architecture of smart gateway with fog computing in which the data is pre-processed and trimmed before sending to server. The network delay could be reduced by applying their architecture. Truong et al. \cite{truong2015software} proposed a new vehicular ad hoc network architecture which combines the software defined network and fog computing. Utilizing fog computing to offer delay-sensitive and location-awareness services could optimize the resources utility and reduce service latency. Zhu et al. \cite{zhu2013improving} proposed a web optimization mechanism in the context of fog computing. In their mechanism, the existing methods for web optimization are combined with the unique knowledge that is only possessed at the fog nodes. The evaluation shows that the web page rendering performance is improved comparing to simply applying the existing methods. Deng et al. \cite{7467406} investigated the tradeoff between transmission delay and power consumption in the fog-cloud computing system. A workload allocation problem targeting the minimal power consumption with the constrained service delay is formulated. They solve the problem with an approximate approach and the solution could enable the fog computing save the communication bandwidth of cloud and reduce transmission latency by sacrificing modest computation resources. To the best of our knowledge, the work proposed in this paper is the first attempt of taking advantage of fog computing to detect faulty sensor in IoT.

\section{Conclusions}
In this paper, an efficient fog-assisted faulty sensor detection scheme with privacy preserving in IoT has been proposed. With the help of fog computing, the faulty sensors could be detected locally and thus saving the communication and computation resource of control center. The Paillier Cryptosystem ensures the sensor data privacy is preserved during the fog aggregation performed by the first layer fog devices. The second layer fog devices could calculate the dispersion measure of smart sensors based on the aggregation results efficiently. The performance analyses have shown that the proposed scheme can achieve a remarkable detection effectiveness and substantially save the computation and communication overhead of control center. The security analysis demonstrates the proposed scheme could achieve authenticity, data integrity and privacy-preservation at the same time. The eavesdropping, man-in-the-middle attack and replay attack from adversary could be completely prevented. In the future, we will investigate other detection schemes for different sensor fault types.

\label{sec8}
\ifCLASSOPTIONcaptionsoff
  \newpage
\fi




\bibliographystyle{IEEEtran}
\bibliography{IoT}
\end{document}